\documentclass[aps,prl,twocolumn,groupedaddress]{revtex4}

\usepackage{graphics} 
\usepackage{graphicx}
\usepackage{dcolumn}
\usepackage{bm}
\usepackage{epstopdf}
\usepackage{amsmath}

\begin{document}

\title{Frequency dispersion of nonlinear response of thin superconducting films in Berezinskii-Kosterlitz-Thouless state.}

\author{Scott Dietrich}
\author{William Mayer}
\author{Sean Byrnes}
\author{Sergey Vitkalov}
\email[Corresponding author: ]{vitkalov@sci.ccny.cuny.edu}
\affiliation{Physics Department, City College of the City University of New York, New York 10031, USA}
\author{A. Sergeev}
\affiliation{SUNY Research Foundation, SUNY at Buffalo, Buffalo, NY14226, USA}
\author{Anthony T. Bollinger}
\author{Ivan Bo\v{z}ovi\'{c}}
\affiliation{Brookhaven National Laboratory, Upton NY 11973, USA}

\date{\today}

\begin{abstract} 

The effects of microwave radiation on transport properties of atomically thin $La_{2-x}Sr_xCuO_4$ films were studied in the 0.1-13 GHz frequency range. Resistance changes induced by microwaves were investigated at different temperatures  near the superconducting transition.  The nonlinear response decreases by several orders of magnitude within a few GHz of a cutoff frequency $\nu_{cut}\approx 2GHz$.  Numerical simulations that assume $ac$ response to follow the $dc$ $V$-$I$ characteristics of the films reproduce well the low frequency behavior, but fail above $\nu_{cut}$. The results indicate that 2D superconductivity is resilient to high-frequency microwave radiation, because vortex-antivortex dissociation is dramatically suppressed in 2D superconducting condensate oscillating at high frequencies.

\end{abstract}
  

\maketitle

Transport properties of thin superconducting films have attracted much interest due to a fascinating physical phenomenon, the Berezinskii-Kosterlitz-Thouless (BKT) transition, predicted to occur at the critical temperature $T_c$ lower than the temperature of the superconducting transition in bulk samples \cite{berezinskii1972,kosterlitz1973,kosterlitz1978,jose2013}. The BKT phase transition originates from long-range (logarithmic) interactions between vortex excitations in a two-dimensional (2D) superconducting condensate. Below $T_c$, the dominant thermal excitations are vortex-antivortex (V-AV) pairs. A superconducting current does not move pairs,  thus producing no energy dissipation. However, the current can break some of V-AV pairs, generate free vortices, set them in motion via the Lorentz force, and thus make the transport dissipative \cite{halperinnelson1979,huberman1978,ambegaokar1980}. The induced dissociation of V-AV pairs depends on the current strength and thus results in an extraordinary violation of the Ohm's law. With this motivation, the strong nonlinear response to the electric current in thin superconducting films was investigated extensively \cite{epstein1981,kadinepsteingoldman1983,fioryhebardglaberson1983,goldman2013}. Despite significant progress, important many-body and edge effects in the V-AV pair dissociation are still under debate \cite{gurevich2008,kogan2007}.  Note that the majority of studies of the nonlinear transport in the BKT regime have in fact been done in the $dc$ domain. How condensate oscillations affect the V-AV dissociation is still unclear. Finally, due to strong phase fluctuations the BKT phenomena are significantly  enhanced in superconducting cuprates, especially in heterostructures with a few superconducting copper oxide layers \cite{lemberger2007}. Our recent studies of $dc$  nonlinearities in MBE-grown heterostructures show that the vortex nonlinearity in the low resistive state exceeds the heating nonlinearity by up to four orders in magnitude \cite{bo2013}.  

Here, we present experimental investigations of nonlinear transport properties of atomically thin $La_{2-x}Sr_xCuO_4$ films, over a broad frequency range from $dc$ to 13 GHz. The experiments indicate a dramatic decrease of the nonlinear response at high drive frequencies, suggesting significant reduction  of the  V-AV pair dissociation  in the oscillating superconducting condensate.

The experiments were performed on thin $La_{2-x}Sr_xCuO_4$ (LSCO) films synthesized by Atomic-Layer-by-Layer Molecular Beam Epitaxy, providing precise atomically thin layers \cite{gozar2008,logvenov2009,bollinger2011}. On the extreme level of control, delta-doping in a single $CuO_2$ layer has been demonstrated \cite{logvenov2009}.  Recently a linear $ac$ response of such films in the BKT state has been studied \cite{gasparov2012,bilbro2011}. Present samples have three distinct layers. The top and the bottom layers, each 5 unit cells (UC) thick, are made of  strongly overdoped ($x=$0.41) normal metals. The sample A, with 5 UC thick inner layer of  $La_{1.72}Sr_{0.28}CuO_4$, shows the BKT transition at $T_c \approx$ 7K \cite{dietrich2014}. The sample B with 1.5 UC thick inner layer of  $La_{1.80}Sr_{0.20}CuO_4$ has $T_c \approx$ 5K. Below we study nonlinear response of the BKT state at $T>T_c$.

\begin{figure}[t]
\includegraphics[width=3.3in]{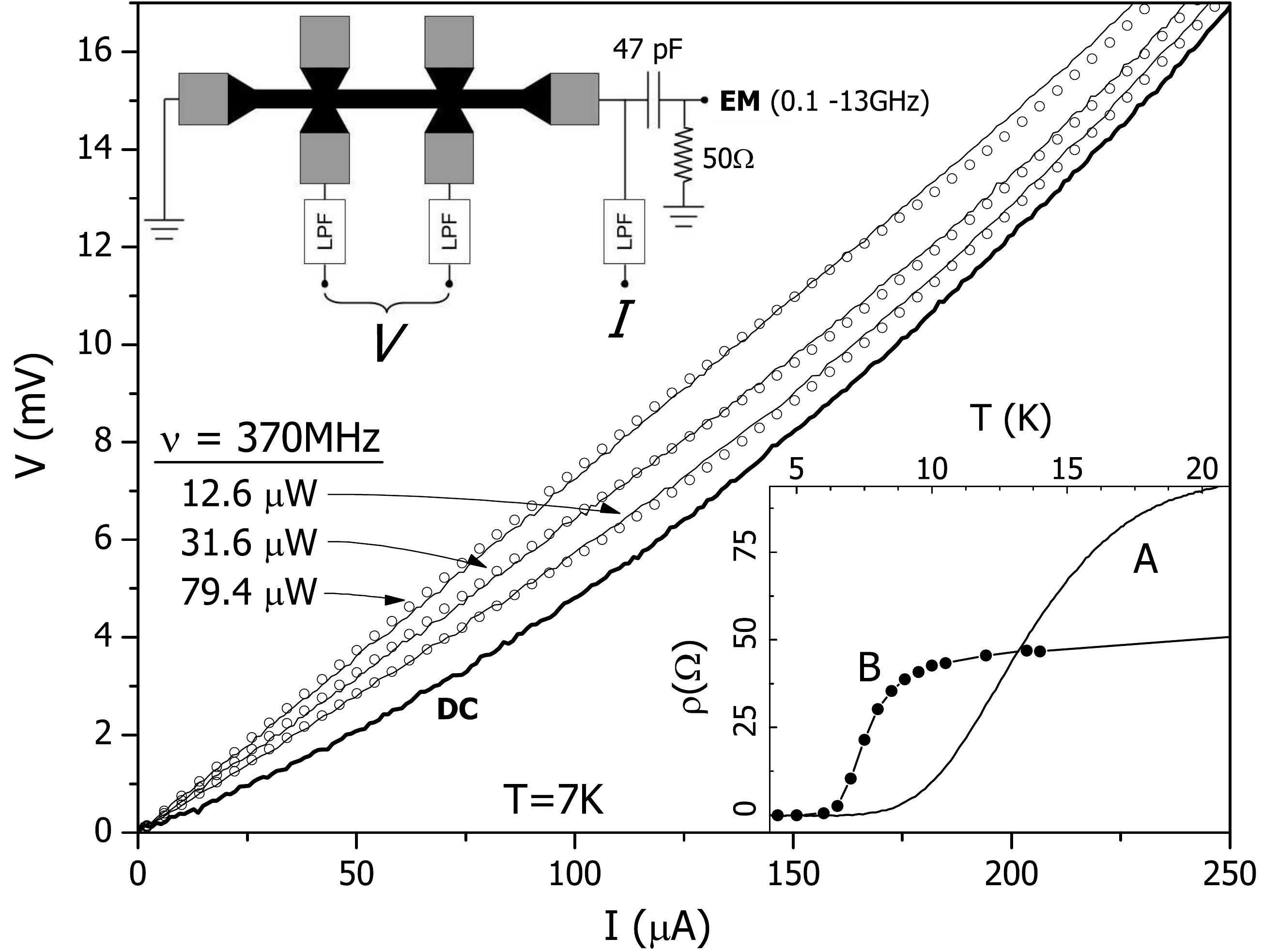}
\caption{Dependence of the voltage $V$ on current $I$ at different EM power applied to 50 $\Omega$ resistor at frequency $\nu=$ 370 MHz. The thick solid line represents the $dc$ response with no radiation applied. The open circles are the results of numerical simulations of the effect of EM radiation on $V$-$I$ curve computed at the same power, for the contact resistance $R_c=$ 210 $\pm$ 30 $\Omega$ in series with the sample B at $T = $ 7 K. The upper inset: a schematic of  $dc$ circuit isolated from EM circuitry by a capacitor $C=$ 47 pF  and three low-pass filters (LPF, $\nu<$ 100 MHz). The lower inset shows sheet resistance per square  as a function of temperature.}
\label{V-I}
\end{figure}

The films were patterned into the shape of Hall-bar devices with the width $W = $200 $\mu$m and the distance between the voltage contacts $L =$ 800 $\mu$m.  A direct current $I$ was applied through a pair of current contacts, and the longitudinal $dc$ voltage $V$ was measured between the potential contacts. The sample and a calibrated thermometer were mounted on a cold copper finger in vacuum. The electromagnetic (EM) radiation was guided by a rigid coaxial line and applied to samples  as shown in the upper insert in Fig. \ref{V-I}. A 50 $\Omega$  resistor terminates the end of the coax and provides  the broadband matching of the EM circuit. The EM power $P$ and amplitude of the microwave voltage $V_\nu$ at the end of the coax were measured {\it in-situ} using the non-selective bolometric response of the 50 $\Omega$  resistor   \cite{dietrich2014}. In what follows, the measured nonlinear response is normalized with respect to  the calibrated EM power $P$, which takes into account all the effects of EM transmission (reflection) to (from) the sample stage.

Fig. \ref{V-I} shows the dependence of the voltage $V$ on the current $I$ taken at different powers $P$ of a low frequency radiation. The thick solid line presents the $V$-$I$ dependence with no radiation applied. The EM radiation increases the resistance as shown by thin solid lines.   To evaluate numerically the radiation effect  the electrical connection (coupling) between the coax and the sample was approximated by  a high-frequency contact resistance $R_c$, which determines the total voltage  $V_S(t)$ and current $I_S(t)$ applied to the sample, assuming that the electromagnetic response  follows the $dc$ nonlinear $V$-$I$ dependence \cite{dietrich2014}. The time averaged $<V_S(t)>$ and  $<I_S(t)>$ are denoted by the open circles in Fig. \ref{V-I}. The resistance $R_c$ was  used  as the single fitting parameter for each computed curve,  providing a  good agreement between the  low frequency experiments and the simulations. 
\begin{figure}[t]
\includegraphics[width=3.4in]{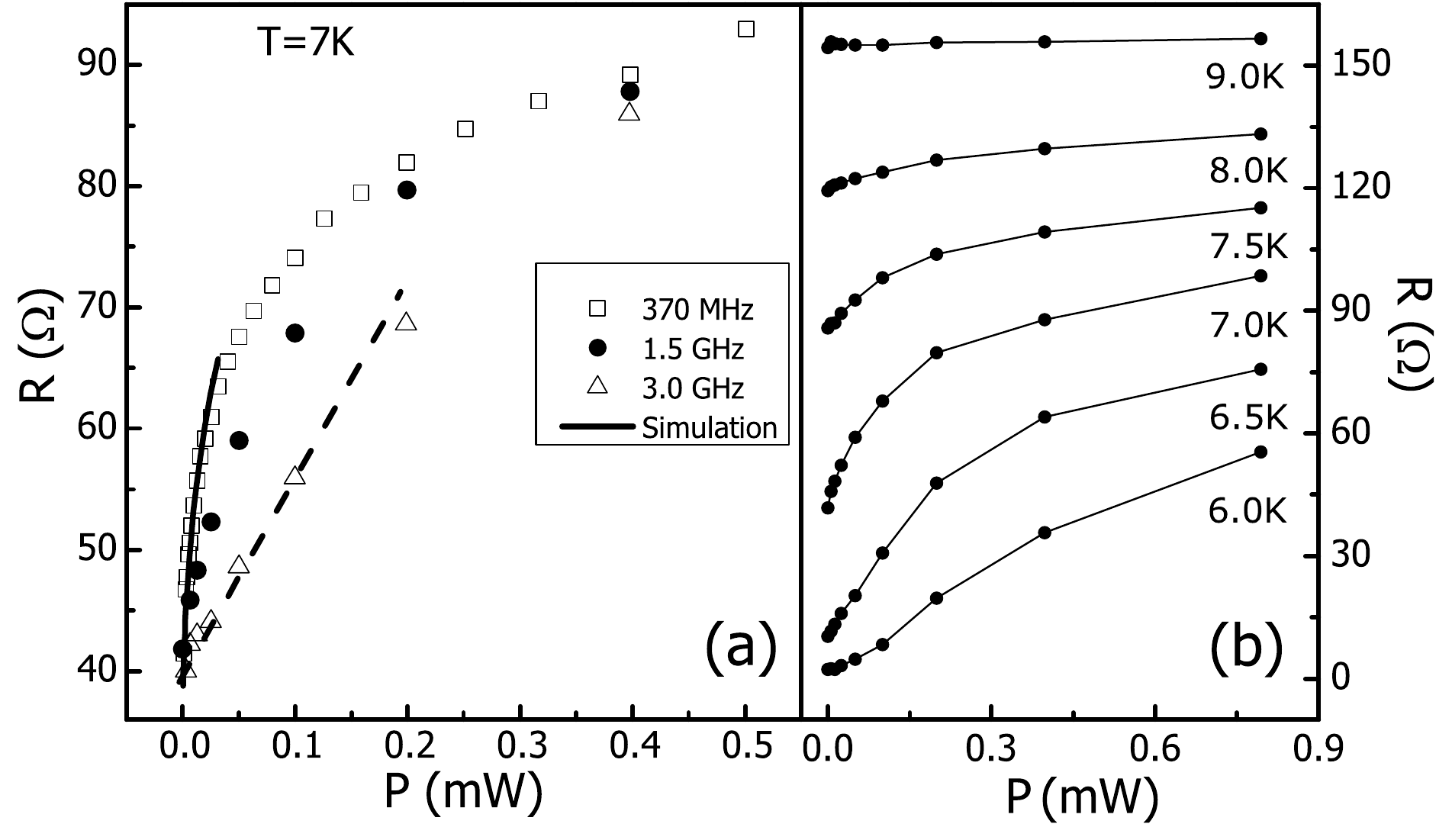}
\caption{(a) Dependence of resistance on radiation power shows strikingly different behavior at low frequency (370 MHz) and high frequencies.  At frequency 3 GHz the power has been linearly scaled down by factor 2 to emphasize a functional  difference in the response to EM radiation. The thick solid line is the numerical simulation  at contact resistance $R_c = 210$ $\Omega$. Dashed line presents  expected dependence  at a small power.  (b) Power dependence of resistance at different temperatures as labeled. $\nu=$ 1.5 GHz. Sample B.}
\label{freqs}
\end{figure}

At small currents, both the experiment and the simulations display a linear relation between the current $I$ and voltage $V$ - Ohm's Law.  Fig. \ref{freqs}b presents the dependence of the ohmic resistance on microwave (MW) power ($\nu=$ 1.5 GHz) taken at different temperatures. The filled dots are the slopes of the $V$-$I$ dependencies at small currents. The power dependence  varies considerably with the temperature. Close to the superconducting state the microwave-induced resistance variations are strong, whereas near the normal state the variations are weak.  The shape of the dependence changes with the temperature. In particular, at $T$ = 6 K the dependence looks more linear than  at $T$ = 6.5 K.

Significant changes of the power dependence   are found in the response to microwaves with  different  frequencies. Fig. \ref{freqs}a shows the power dependencies of the resistance obtained  at frequencies $\nu$ = 0.37, 1.5, and 3 GHz. At low frequency $\nu$ = 0.37 GHz, the dependence  is in a good agreement with the numerical simulation. At high frequencies, the nonlinear response is  much weaker and  has  a different  functional form.   To highlight the  difference,  the values of MW power at frequency 3 GHz  were scaled down by a factor of 2.  A comparison of power dependencies indicates  that at a low power  the  nonlinear response  is considerably weaker  at a high frequency (3 GHz) than  at a low frequency (0.37 GHz). At a high power the strength of the high frequency nonlinearity is  restored. 

\begin{figure}[t]
\includegraphics[width=3.4in]{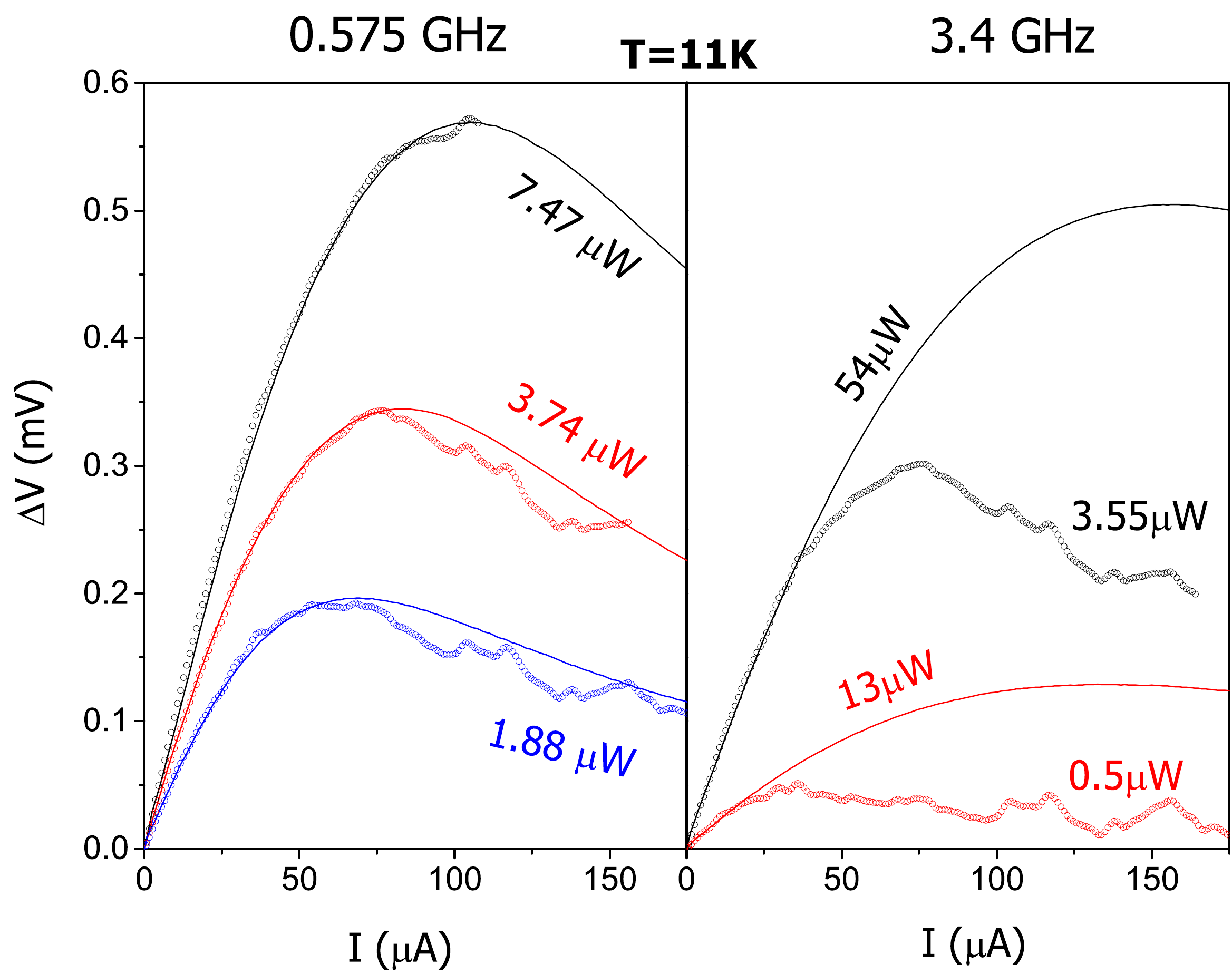}
\caption{(Color online). The change of voltage $\Delta V=V(P)-V(0)$ due to applied EM radiation  as a function of the $dc$ bias $I$ at different microwave powers as labeled. The solid lines are experimental data; the open circles present numerical simulations at $R_c=63 \pm 2\Omega$. Comparison of the left and right panels shows the agreement between experiment and simulation at low frequencies, but serious disparity at high frequencies. Sample A.}
\label{deltaV}
\end{figure}

Fig. \ref{deltaV} shows the dependence of radiation-induced variations of the $dc$ voltage $\Delta V$  on  $dc$ bias $I$ taken at different frequencies and power levels, as labeled. At small currents, the response is linear, indicating no observable $ac$ rectification in the device.  The left panel demonstrates the effect of the low-frequency radiation ($\nu$ = 0.575 GHz) on the resistance. All three curves are in very good agreement with the ones obtained by numerical simulations for the same radiation powers \cite{dietrich2014}.  The simulations replicate all the details of the experiments including the shift of the observed maximum with increased  power using single fitting parameter $R_c=63 \pm 2$  $\Omega$. The right panel shows the effect of the high frequency ($\nu$ = 3.4 GHz) radiation. One can see that the high-frequency response does not follow the $dc$ $V$-$I$ curve and cannot be explained by a reduction  (re-scaling) of the  MW current.

\begin{figure}[t]
\includegraphics[width=3.2in]{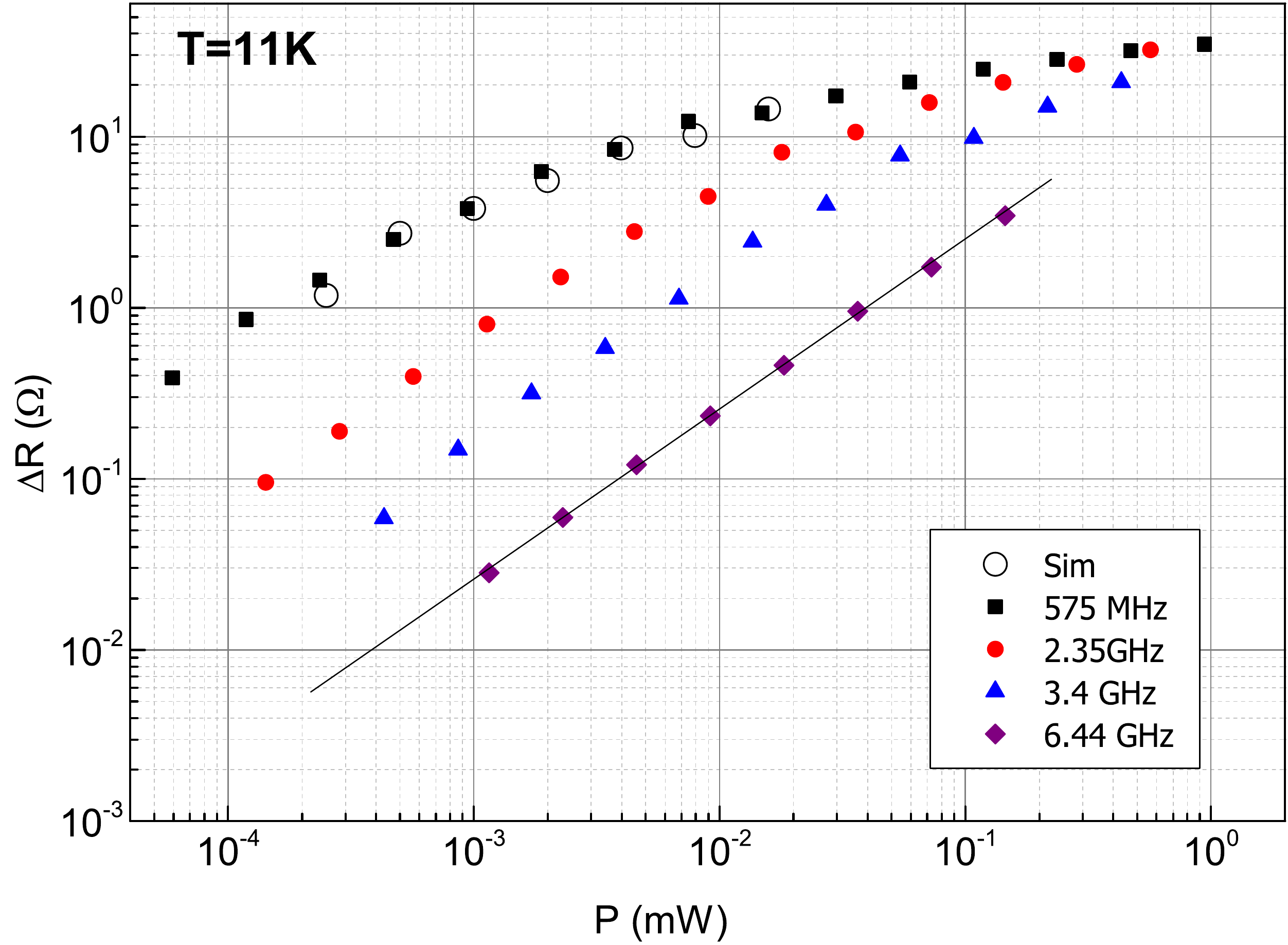}
\caption{(Color online) The power dependence of  radiation induced resistance variations $\Delta R=R(P)-R(0)=\Delta V/I$ obtained at small $dc$ biases  $\Delta V \sim I$  (see Fig. \ref{deltaV}).   The solid line is the dependence $\Delta R \sim P$ expected for small power $P$. Open circles are numerical simulations at $R_c=$ 62 $\Omega$. Sample A.}
\label{slopes}
\end{figure}

At small currents, the linear part of the voltage variations $\Delta V(I) \sim I$ is related to the change of the ohmic resistance: $\Delta R=R(P)-R(0)=\Delta V(I)/I$.     Figure \ref{slopes} displays the behavior of the resistance variation $\Delta R$ with the power at different frequencies.  At a small power, the  induced resistance variations  are proportional to the power. At higher powers, the dependence is weaker. At higher frequencies, the transition to weak power dependence occurs at a higher power. At the highest power levels all dependencies converge (see Fig. \ref{freqs}a). 

An analysis indicates several distinct features of the $dc$ nonlinear response presented in Fig. \ref{V-I}. At small currents the $V$ - $I$ dependence is well approximated by a combination of linear and cubic terms \cite{bo2013}. The dependence is presented below: 
\begin{equation}
V=R_0\cdot I+\gamma I^3 ,
\label{main}
\end{equation}
where $R_0$ is the Ohmic resistance, the coefficient $\gamma$ is a constant. The high-current behavior is in agreement with the one expected within the BKT scenario: $V \sim I^\alpha$ \cite{halperinnelson1979,kadinepsteingoldman1983,fioryhebardglaberson1983}. The exponent $\alpha(T)$ decreases from 8 to 1 as the temperature increases, indicating a BKT transition at $\alpha(T_c)=3$ \cite{dietrich2014}.    In accordance with Eq.(\ref{main}) at small  $ac$ currents $I_\nu(t)$, the voltage variations  $\Delta V(I) =<V(I+I_\nu(t))>-V(I) \approx 3\cdot I \cdot <I_\nu(t)^2>$ are proportional to the $dc$ bias $I$, which agrees with Fig. \ref{deltaV}, and proportional to the RF power $P \sim I_\nu^2$, which agrees with Fig. \ref{slopes}. The decrease of the nonlinear response shown in Fig. \ref{deltaV} at high $dc$ biases and the transition to a weaker power dependence presented in Fig. \ref{slopes} are the results of the high current response   $V \sim  I^\alpha$   at $\alpha<$2. 

\begin{figure}[t]
\includegraphics[width=3.4in]{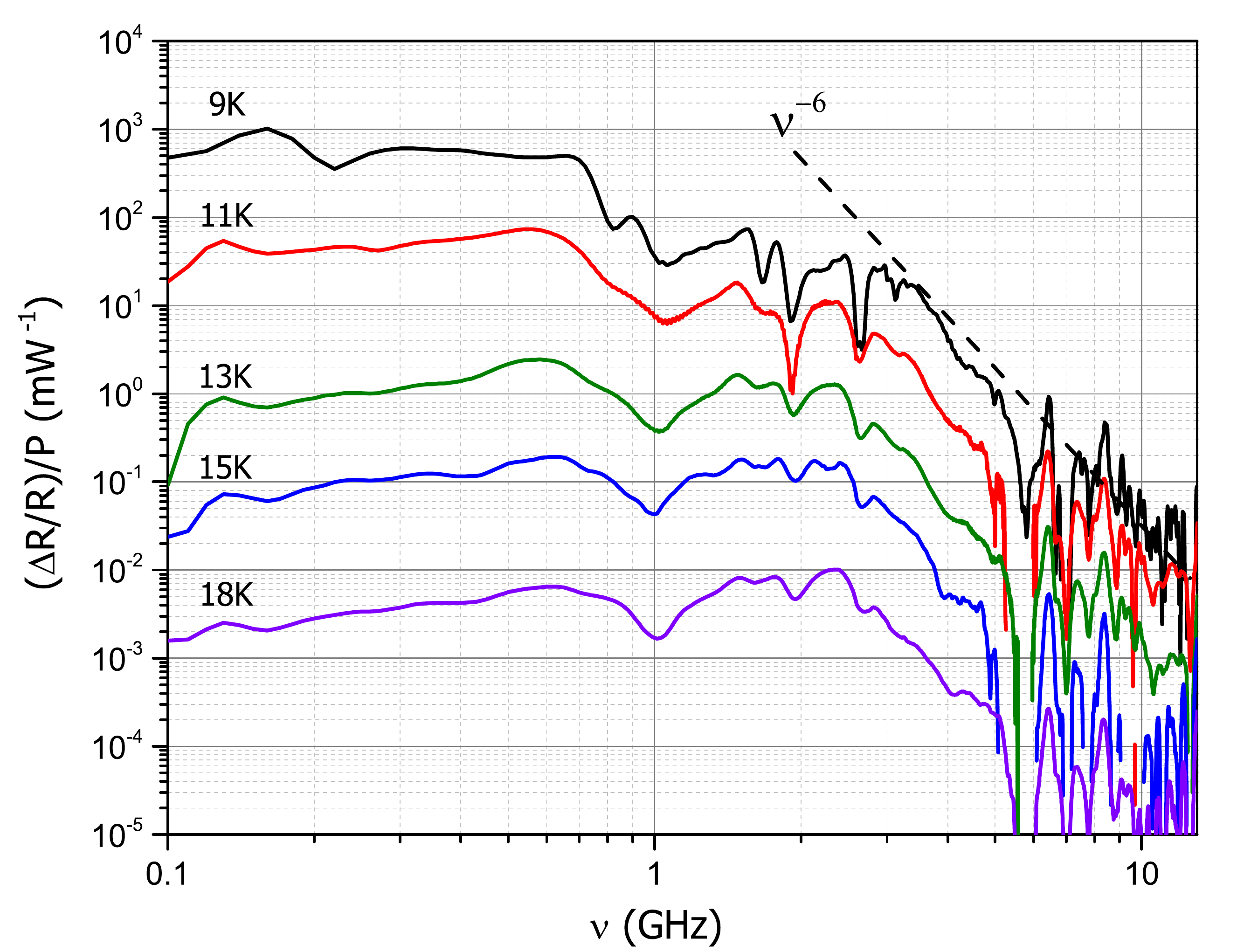}
\caption{Frequency dependence of the relative variations of the resistance $\Delta R/R$ normalized by applied  power $P$ at different temperatures as labeled. Dashed line is an approximation  of the nonlinear response between 3-6 GHz. Sample  A.}
\label{diff_F}
\end{figure}

Fig. \ref{diff_F} presents the frequency dependence of the nonlinear response, taken at different temperatures and small powers $P$.  At low temperatures, 9 K $\le T \le $11 K, the response  decreases by 3-4 orders in magnitude, as the frequency increases above few GHz.  This is in a good agreement with the reduction of the $dc$ nonlinearity by 3-4 order of magnitude observed between the regimes of V-AV depairing at 10 K and electron heating in the normal state (see Fig. 3b in \cite{bo2013}). 

To be sure that the observed effect  is not related to a strong decrease in the microwave coupling, we have evaluated the MW current through samples by investigating the MW reflection in the same setup \cite{dietrich2014}. These studies indicate some frequency dispersion in the MW current.  However, the  dispersion is nearly uniform and significantly smaller (a variation by a factor of 3-4) than the observed  reduction of the nonlinear response by 3-4 orders of magnitude. Independent measurements of MW voltage, current, and the nonlinear response in several samples allow us to conclude that the observed reduction of the nonlinearity  is of fundamental nature.    

Since the reduction is strong, we associate it with the frequency suppression of the dominant mechanism  in BKT regime - the current induced V-AV pair dissociation.  If in the course of high-frequency oscillation the distance between the vortex and antivortex within a pair does not exceed a critical distance $l_c$, then the pair can survive.   At higher power, the amplitude of  V-AV oscillations may exceed the critical distance $l_c$, making the response similar to the one obtained at low frequency. It corresponds to  Fig. \ref{freqs}a and Fig. \ref{slopes}.

An analysis of the oscillating vortex motion inside a V-AV pair in the presence of $dc$ depairing current shows rich physics, analogous to the Kapitza pendulum with a vibrating pivot point \cite{landau1976}.  Position $x_v$ of a vortex with the mass $M$ moving in a medium with the viscosity $\eta$ under effect of a time-dependent external potential  $U(r,t)$ is given by  
\begin{equation}
M\frac{d^2 x_v}{dt^2}+\eta\frac{d x_v}{dt}  =-\frac{\partial U(r,t)}{\partial r},
\label{vortex}
\end{equation}
The potential $U$ is a sum of V-AV interaction and a potential of Lorentz force \cite{ambegaokar1980}: $U(r,t)=2q_0^2 ln(r/\xi)-q_0J_s(t) r$, where $\xi$ is a size of the vortex core and  $q_0^2=\pi n_s \hbar^2/4m$, where $n_s$ and $2m$ are density and the mass of superconducting carriers. We express the supercurrent density  $J_s=J_{dc}+J_\omega$ as a sum of the $dc$ bias and  current  oscillations $J_\omega$ at angular frequency $\omega=2\pi \nu$.  An analysis of $ac$ vortex motion for critical pairs near the saddle point of the potential ($\partial U/\partial r \approx 0$) shows a reduction of the vortex displacement at high frequencies: $x_v^{ac} \approx q_0J_\omega/(-M\omega^2+i\omega \eta)$.  Moreover, at $\omega > \omega_c=\eta/M$, the $ac$ depairing is strongly suppressed due to the phase shift between the vortex displacement and the barrier height. At the moments when the potential $U(t)$ reaches a minimum, the displacement is the smallest, thus preventing the V-AV dissociation. Finally at high frequencies a nonlinear analysis \cite{landau1976} of Eq.(\ref{vortex})  indicates an  effective  attractive force inside the pair, which is proportional to the square of the vortex  displacements: $\delta F_{eff} \sim \frac{\partial^3 U}{\partial r^3} <(x_v^{ac}(t))^2>\sim \frac{<(x_v^{ac})^2>}{r^3}$. These effects reduce the pair dissociation at high driving frequencies.   Although the origin of the cutoff frequency $\nu_{cut}$ requires  further research, we  note that  an evaluation of  the frequency $\nu_c = \eta/(2\pi M)$ is in a good agreement with our data. Using  the Stephen-Bardeen viscosity \cite{stephen_bardeen1965}   and  considering the vortex mass $M$ as a total mass of electrons in the core \cite{volovik1997} we get the characteristic frequency of $\nu_c$ = 2 GHz at the core radius of 8 nm, which is approximately two times bigger than the superconducting coherence length $\xi$. The above analysis points toward the crucial importance of the displacement-force phase relations in the nonlinear $ac$ response of the bulk BKT state supporting substantially the fundamental origin of the observed phenomenon.   
                                                                                                                                                                                                                                                                                                                      
In summary, strong nonlinear response  to low frequency radiation is observed in atomically thin superconducting films,  in BKT state.  The response decreases by several decades  within a few GHz above the cutoff frequency $\nu_{cut} \approx$ 2 GHz. This indicates  that 2D superconductivity is quite resilent  to the high frequency radiation because of  a strong reduction of the vortex-antivortex dissociation in  oscillating 2D superconducting systems. This general conclusion is in agreement with the results of the linear response studies of the BKT state in thin disordered films of traditional superconductors. In particular, detailed investigations of the conductivity in the range 9-120 GHz show only frequency-dependent Drude absorption without any measurable dissipation due to vortex-antivortex dissociation \cite{crane2007}. 

Sample synthesis by ALL-MBE and characterization (I.B.) and device fabrication (A.T.B.) were supported by U.S. Department of Energy, Basic Energy Sciences, Materials Sciences and Engineering Division. Work at CCNY and SUNY was supported by National Science Foundation, Division of Electrical, Communications and Cyber Systems (ECCS 1128459).

\section{Supplemental Material.}
\section{Transport in DC Domain. BKT transition.}

\begin{figure}[b]
\includegraphics[width=3.2in]{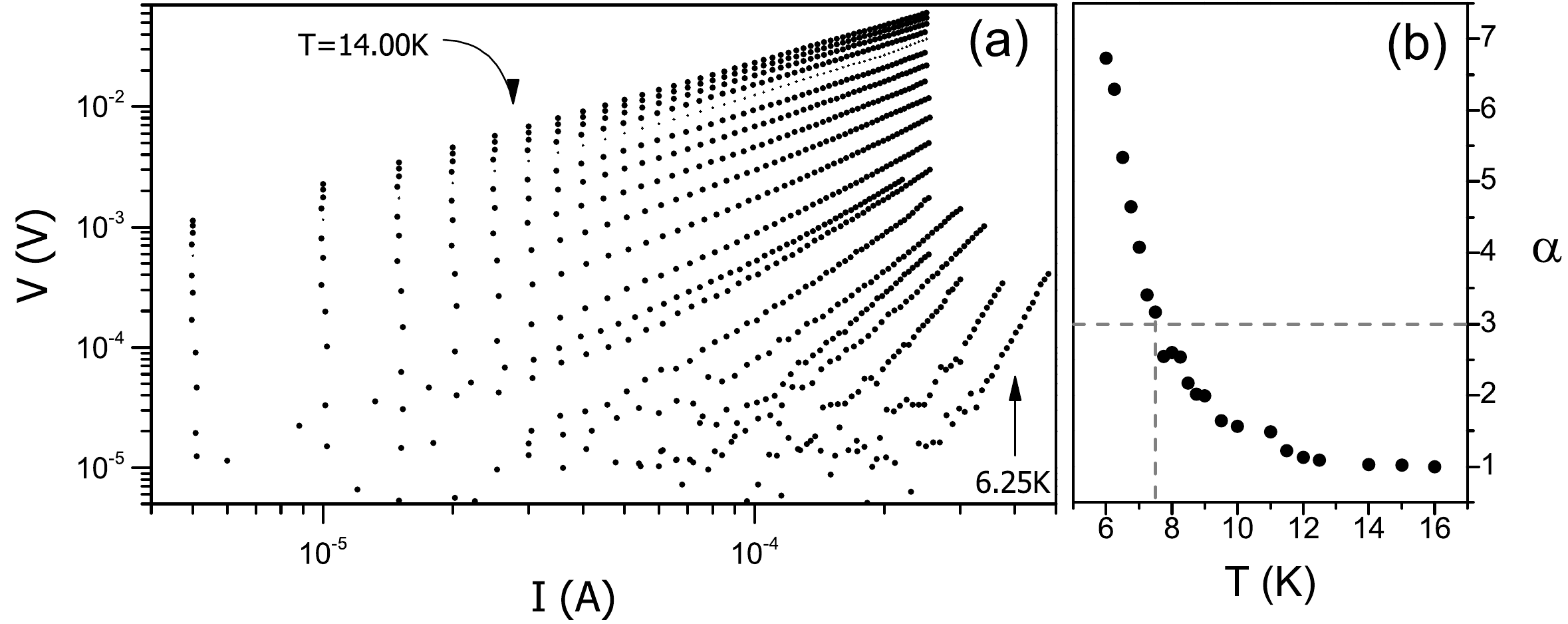}
\caption{(a) Shown here are the $dc$ I-V characteristics for Sample A to which a $V\sim I^\alpha$ fitting has been done at high currents, where the curves display a clear power-law form even at low temperatures. (b) Analysis of the temperature-dependent exponent $\alpha(T)$ demonstrates a BKT transition at $T\approx$ 7-8K.}
\label{IVs}
\end{figure}

 According to the BKT theory, $V$-$I$ characteristics should display a $V\sim I^{\alpha}$ dependence where the temperature-dependent exponent, $\alpha(T)$, ranges from one in the normal state and increases as temperure decreases, passing through it's critical value $\alpha(T_c)=3$ at the Brezenskii--Kosterlitz--Thouless (BKT) transition temperature, $T_c$ \cite{halperinnelson1979,kadinepsteingoldman1983,fioryhebardglaberson1983}. Figure \ref{IVs}a presents $V$-$I$ dependencies of Sample A plotted in log-log scales. The dependencies obtained at different temperatures  in a range from 6.25-14.00 Kelvin at zero  magnetic field. At high applied currents the dependencies are in agreement with the power law $V\sim I^\alpha$, where coefficient $\alpha$ depends on temperature.

Fitting these curves to the power-law $I$-$V$ function yields the temperature dependence of $\alpha(T)$. Figure \ref{IVs}b presents the results of the fitting. The coefficient $\alpha$ varies from about 7 at 6.25K to 1 at at 12K  indicating  a BKT transition at $T_c\approx$7.5 K corresponding to  $\alpha=3$.  We note that in accordandce with the thermal conductance per unit area of the sample  $g_{e-ph}\approx 1\cdot(T/20)^4$ W/Kcm$^2$ \cite{bo2013} the electron overheating  at temperatures  $T$=7-10K  near the BKT transition  is below 0.1 K at the highest currents applied to the sample. Thus contributions of  thermal effects to the observed nonlinearity are negligibly small.

\section{Numerical Simulations and Bolometric Calibration.}

\begin{figure}[b]
\includegraphics[width=3.2in]{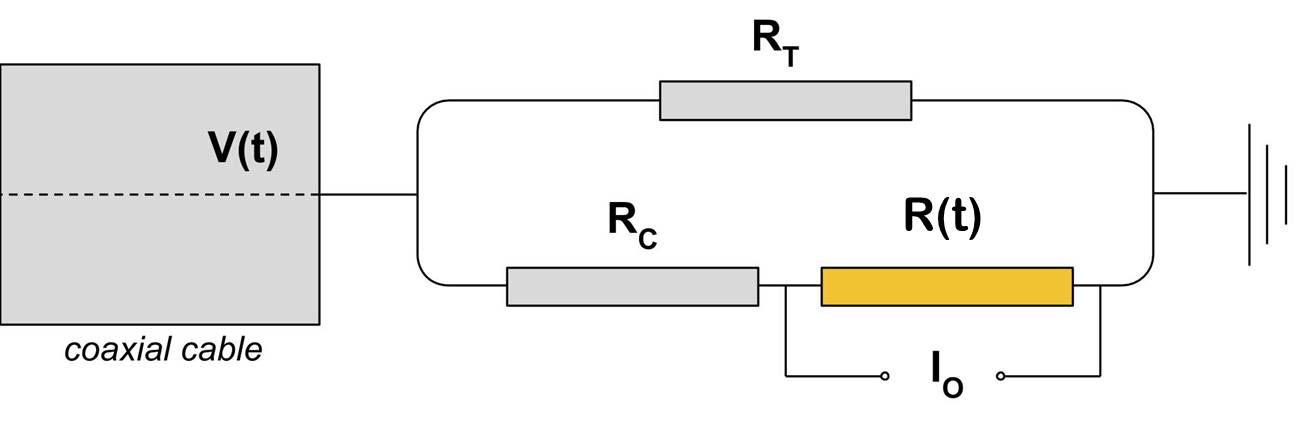}
\caption{(color online)  Shown here is the microwave circuit, which is used in the simulation  to obtain the voltage and current in the sample. The circuit contains a terminal resistor ($R_T$), contact resistance ($R_c$, a fitting parameter), and the sample itself ($R(t)$, yellow). A $dc$ bias $I_0$ is applied  to the sample through dedicated leads equipped with low pass filters(not shown). In addition the $dc$ and microwave circuits are  separated by 47 pF  capacitor, which has been included in $R_c$.  } 
\label{simdiagram}
\end{figure}

The goal of the simulation is to use  $dc$ I-V characteristic of the sample to reproduce the I-V characteristics under  the MW excitation. Figure \ref{simdiagram} presents the approximation of the microwave circuit used in experiments. The circuit contains a 50 Ohms terminal resistance $R_T$  which provides a broadband coupling of the circuit with microwaves.   The nonlinear resistance $R(t)$ describes  the sample response. The contact resistance $R_c$ couples microwaves with the sample.To isolate the $dc$ and the microwave circuits a 47 pF capacitor is added in series with the sample.  The AC resistance of the capacitor is included in the $R_c$.    

The simulation is based on the assumption that the nonlinear response to EM radiation follows the $dc$  I-V characteristics. The response of the circuit is described by the following set of equations:
\begin{subequations}
\begin{align}
& V(t)=I_SR_c+V_S\\
&I(t)=I_0+I_S(t)\\
&V_{0}+V_S(t)=V_{DC}(I(t))
\end{align}
\label{sim}
\end{subequations}

Eq.(\ref{sim}a) describes voltage in the microwave (AC) circuit. The AC voltage $V(t)$ at the end of the coax is the sum of the voltage applied to  the contact resistance $I_SR_c$ and the AC voltage applied to the sample $V_S$, where  $I_S$ is the AC current through the sample and the resistance $R_c$. Eq.(\ref{sim}b) describe the total current $I(t)$ through the sample, which is sum of AC ($I_S$) and $dc$  ($I_0$) currents. The third equation relates the voltage applied to the sample to the total current, using the $dc$   V-I characteristic obtained in the experiment.           

The equations yield  AC current $I_S(t)$ and voltage $V_S(t)$ at  given microwave voltage  $V(t)=V_\omega cos(\omega t)$,  contact resistance $R_c$ and $dc$  current $I_0$. The time average of the AC response yields contributions to $dc$  current $\delta I_0=<I_S(t)>$ and voltage $\delta V_0=<V_S(t)>$ and makes a  V-I dependence in the presence of the microwave radiation.  The V-I characteristic depends on  the contact resistance $R_c$. A comparison between  simulated and the actual V-I characteristics yields the contact resistance $R_c$ at given MW voltage (power $P_{sim}$).  The tables below give the fitting parameter $R_c$ at different powers $P_{sim}$.

\begin{center}

\begin{tabular}{ l l}

Figure 1.&
\begin{tabular}{| c | c | c | }
	\hline
 	$P_{sim} (\mu W)$ & $R_c (\Omega)$ \\ \hline
    	79.4  & 240   \\ \hline
    	31.6 & 200   \\ \hline
    	12.6 & 180  \\
	\hline
 \end{tabular} \\
&\\
Figure 2. &
\begin{tabular}{| c | c | }
	\hline
 	$P_{sim} (dbm)$ & $R_c (\Omega)$\\ \hline
    	-40 to -15  & 200 \\ 
	\hline
 \end{tabular} \\
&\\
Figure 3. &
\begin{tabular}{| c | c | c | }
	\hline
 	$P_{sim} (\mu W)$ & $R_c (\Omega)$ &  $P_{exp} (\mu W)$\\ \hline
    	1.88  & 61 & 1.88 \\ \hline
    	3.74 & 61 &  3.74 \\ \hline
    	7.47 & 61 & 7.47\\ \hline
	0.5 & 65 &  13.0 \\ \hline
	3.55 & 65 &  54.0 \\ \hline
 \end{tabular}\\
\end{tabular}
\end{center}

We note that  for all Figures except Fig.3 $P_{sim}=P_{exp}$, where $P_{exp}$ is a calibrated power at the end of the coaxial line. 

Fig. 3 shows a reduction of the upper limit of the $dc$   bias for the simulations at high MW powers. This  is related to the limited range of currents in the experimental $dc$   V-I dependences used for the numerical simulation: $[-I_{max}, I_{max}]$. In accordance with Eqs.(\ref{sim}b,c)the simulation calculates the voltage $V_0+V_S = V_{DC}(I_0+I_S(t))$ using the $dc$   V-I dependence $V_{DC}(I)$. To obtain the voltage V the value $I_0+I_S$ must be less $I_{max}$. In Fig.3 it reduces progressively the range of the $dc$   biases $I_0$ at which the simulation works at higher MW currents (powers).  

\subsection{Bolometric calibration of MW power.}
\begin{figure}[t]
\includegraphics[width=2.8in]{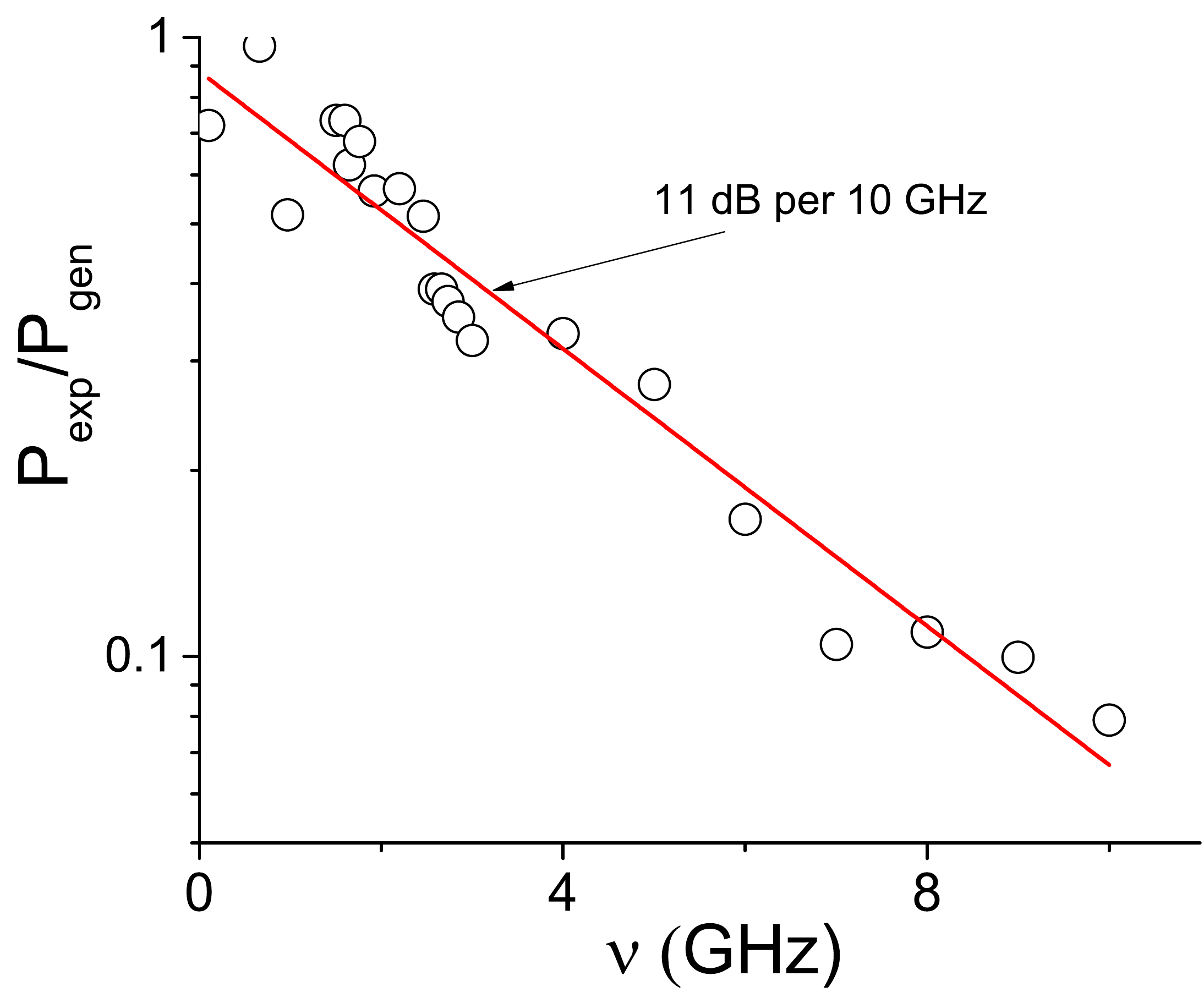}
\caption{Ratio between power $P_{exp}$ supplied to  the end of the coaxial line at low temperatures and the  calibrated power $P_{gen}$ at the output of microwave generator vs frequency. The straight line is the approximation used to normalize the nonlinear response shown in Fig. 5 of the main text.}
\label{bolometric}
\end{figure}

The {\it in-situ} bolometric calibration of the microwave power  applied to the end of the coax is based on the following measurement. In experiment the resistor $R_T=50 \Omega$ terminating the coaxial line (see Fig.\ref{simdiagram}) is thermally anchored to a copper sample holder. A  microwave power $P_{exp}$, applied to  the  end of the coax, heats the resistor $R_T$.  It decreases   the amount of the $dc$ electric power, $P_h$, supplied by a temperature control unit to  $dc$ heater, which is  attached to the sample holder to maintain the sample temperature. The change of the $dc$ heating power $\Delta P_h=\Delta (V_{dc}^2/R_h)$ is found by measuring voltage $V_{dc}$ applied to the heater (resistance $R_h$=100 $\Omega$) with and without MW radiation. The bolometric calibration of the MW power $P_{exp}$ is based on the conservation of the total energy supplied to the stage: $\Delta P_h=P_{exp}$.  

Fig.\ref{bolometric} shows the ratio of the calibrated power $P_{exp}$ to the power $P_{gen}$ at the output of the MW generator at different frequencies.   The  calibrated power $P_{exp}$ decreases by about 10 times at high frequencies. The power decrease is  found to be in agreement (within 40\% deviations) with  total microwave losses in coaxial lines between the MW generator and the sample, which were measured separately in transmission and reflection experiments. This agreement verifies the expected independence of  the power losses in the terminal resistance  $R_T$ on the frequency.  

The bolometric calibration is done in the presence of the superconducting samples.  The fact that the decrease in supplied MW power is largely due to MW losses in transmission lines indicates also a relatively weak effect of the superconducting samples attached to the end of the coaxial line  on the overall microwave reflection from the sample stage. It is in accord with measurements of the reflection coefficient from the sample stage presented in the next section. The bolometric calibration allows computing the microwave voltage at the end of the coax: $V_\omega=(P_{exp}\cdot R_T)^{1/2}$. The voltage  is  used in the numerical simulations.

\section{Microwave Coupling to the Samples: Reflection Measurements.}

\begin{figure}[t]
\includegraphics[width=3.2in]{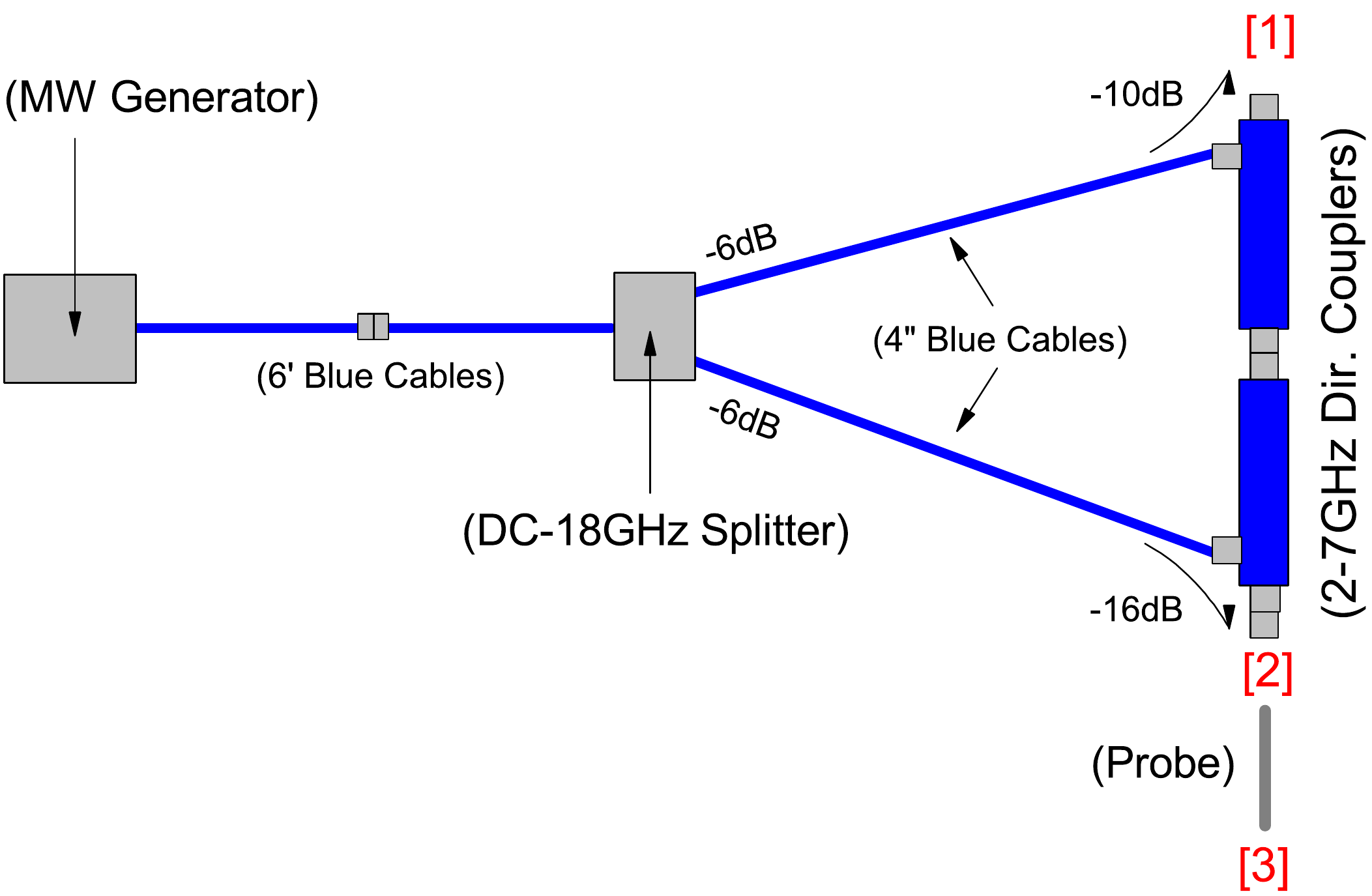}
\caption{Schematic for microwave reflection measurements with signals in the range of 1-7 GHz. The broadband splitter splits microwaves into two channels.  The reference signal goes through a directional coupler to a MW detector attached to port [1].  The incident signal goes down through another directional coupler and  the probe, attached at  port [2],   to a sample stage attached to port [3]. The incident signal is reflected  back  through the probe to the  MW detector (port [1]), where the interference occurs between  the reflected and reference signals.}
\label{setup}
\end{figure}

In the low frequency domain the nonlinear behavior (and the contact resistance $R_c$) is found to be nearly frequency independent taking into account the bolometric calibration of the power applied to the end of the coaxial line.  At high frequencies the response does not follow the $I-V$ curves and the extraction of the contact resistance from the comparison with the numerical simulations is not possible.  To evaluate the coupling of the EM radiation to the sample at high frequencies broad band measurements of  a reflected microwave power   were performed at two temperatures $T_n$ and $T_s$ corresponding to normal and superconducting states of the samples. Since in the normal state the superconductivity is absent the  difference between two reflected signals indicate the strength of the coupling of the MW radiation to the superconducting condensate at different frequencies. The microwave broad band measurements (1 to 7 GHz)  demonstrate that microwaves are quite uniformly coupled to the superconducting condensate in the studied frequency range indicating that  frequency variations of the contact resistance $R_c$ are not significant for the observed strong reduction  of the nonlinear response at high drive frequencies. Details of these experiments are presented below.

The setup of the reflection experiments is shown in Fig.\ref{setup}. The microwave power supplied by a microwave generator splits between two channels. The reference (upper) channel provides  a reference signal $V_c$  supplied through a broad band directional coupler (-10dB) to a broad band microwave detector, which is attached to port  [1]. The sample (lower) channel supplies microwave radiation to another directional coupler (-16dB), which directs the radiation to the sample attached to port [3] of a semi-rigid coaxial line (probe) connected  to the port [2] of the coupler.  Reflected  from the sample  signal $V_r$ is guided by the same coax back to the detector, where it interferes with the reference signal $V_c$. Due to a significant difference in lengths of the reference and sample channels variations of the microwave wavelength (frequency) yield oscillating interference signal on the detector.  At a small microwave amplitude the detected signal $V_{det}$ is proportional to microwave power $P$, in other words, to the square of the microwave voltage:

\begin{figure}[t]
\includegraphics[width=3.2in]{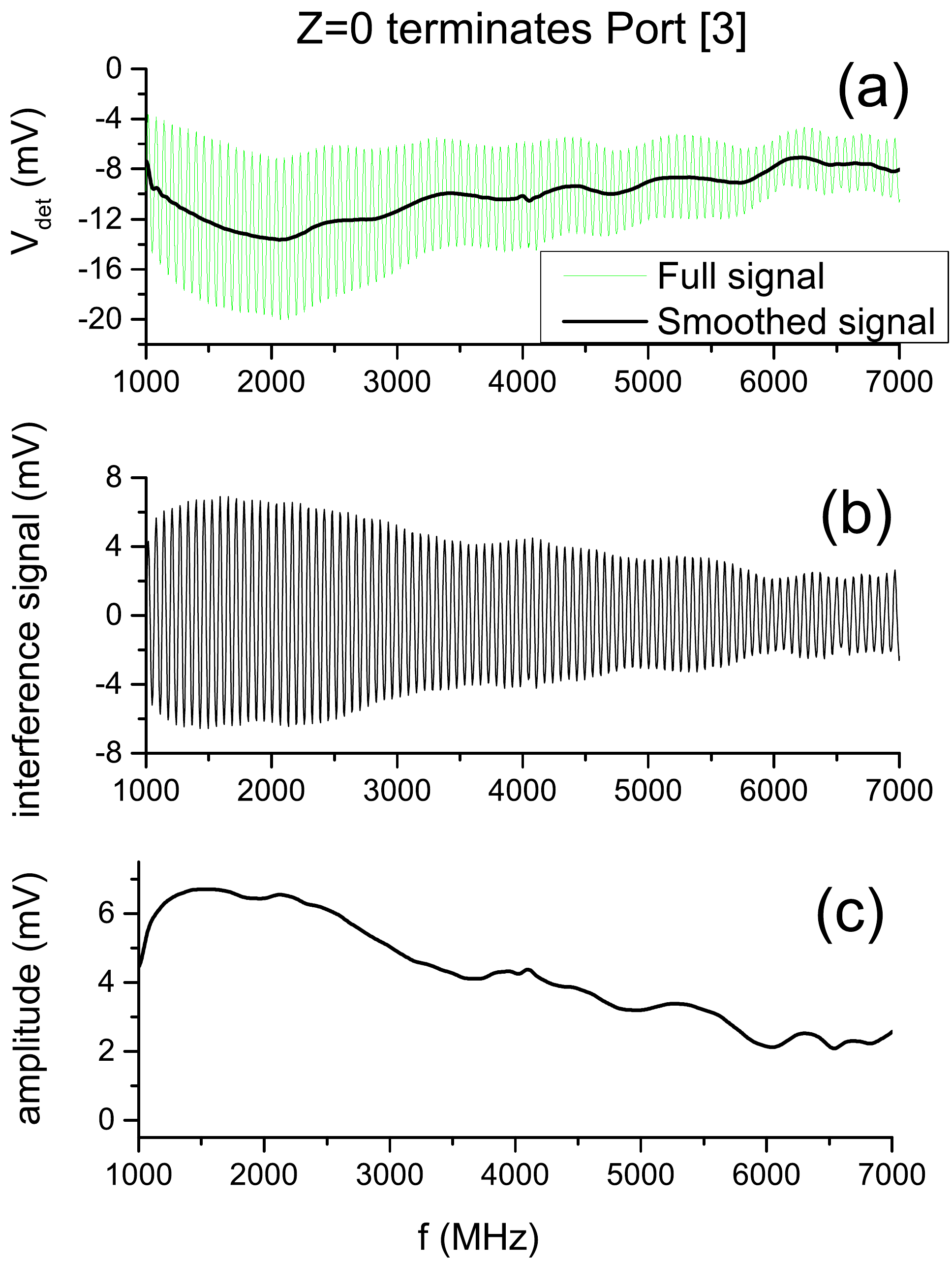}
\caption{The above graphs demonstrate the procedure  for extracting the amplitude of the interference term between signal reflected from port [3] and the reference signal $V_c$. The figure presents the signal reflected from port [3] terminated by zero impedance ($Z=0$). The same procedure is used for the signal reflected by the sample. (a) The full signal (green) is smoothed using a FFT filter (black). (b) Subtracting the smoothed signal from the full signal leaves only the interference term with frequency dependent amplitude shown in (c).}
\label{Z=0}
\end{figure}

\begin{subequations}
\begin{align}
V_{det}& \sim (\vec V_c+\vec V_r)^2\\
&\sim V_c^2+2 \cdot V_c \cdot V_r \cdot \cos{\phi}+V_r^2\\
&\sim V_c^2+2 \cdot V_c \cdot \gamma^2\Gamma V_{in} \cdot \cos{\phi}+V_r^2
\end{align}
\end{subequations}
where these $\vec V_c$ and $\vec V_r$ are the reference and reflected voltages, considered on the phasor diagram  and  $\phi$ is the phase difference between $V_c$ and $V_r$ signals. In the last equation the reflected signal $V_r=\gamma^2\Gamma V_{in}$ was substituted with the incident voltage $V_{in}$ at port [2], using  the reflection coefficient $\Gamma$ at the bottom of the probe (port [3]) and  the  microwave losses $\gamma$ in the coaxial line between ports [2] and [3].  

Terminating the port [3] by zero resistance (terminal impedance Z=0) one can measure the interference pattern corresponding to the complete reflection of microwaves  from the port [3].   Figure \ref{Z=0}(a) presents the detector signal $V_{det}$ corresponding to  the complete microwave reflection (Z=0) as function of microwave frequency. Fast oscillations corresponds  to the interference between $V_c$ and $V_r$, while the slow variations of the signal corresponds mostly to variations of the reference signal $V_c$. Contributions of the reflected signal $V_r^2$ to the microwave background are less than 10\% and are neglected below. To separate the interference term (fast oscillations) from the background a FFT filter was applied resulting  in the thick curve presented in Fig.\ref{Z=0}(a). A subtraction of  the background from the detector response yields the interference term                 
$2 \cdot V_c \cdot \gamma^2\Gamma V_{in} \cdot \cos{\phi}$ shown in Fig.\ref{Z=0}(b). In the case of the complete reflection the reflection coefficient $\Gamma_0=-1$. The corresponding amplitude of the interference term $2  \gamma^2\vert \Gamma_0\vert V_cV_{in}$  is shown in Fig.\ref{Z=0}(c). A decrease of the interference  amplitude below 2 GHz is related to  the decrease of $V_c$ and $V_i$ due to a properties of the couplers, which are designed to work in the range between 2 and 7 GHz.  Decrease of the amplitude at high frequencies is related to microwave losses in the coaxial cable ($\gamma^2$).

\begin{figure}[t]
\includegraphics[width=3.2in]{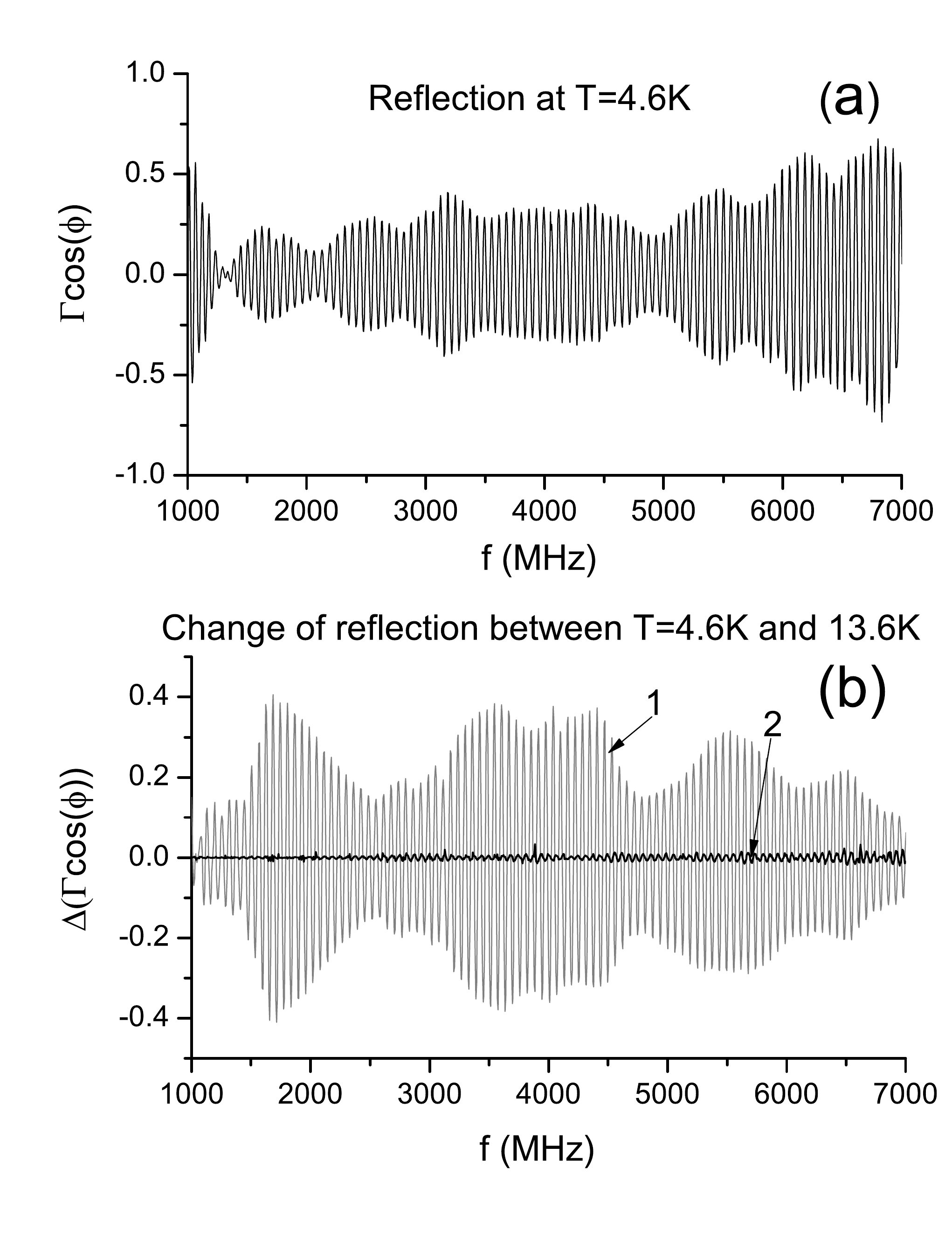}
\caption{(a) The frequency dependence of the reflection coefficient for Sample C at $T=$4.6K is found by dividing the interference signal from the sample by the amplitude of the interference signal corresponding to the perfect reflection ($Z=0$ terminal resistance shown in Figure 4c). (b) Curve 1 presents the change in the reflection coefficient from the sample stage attached to port [3], when the sample undergoes transition from the normal ($T=$13.6K) to superconducting ($T=$4.6K) states. Curve 2 presents the change of the reflection from the stage without the sample in the same temperature interval.}
\label{reflection}
\end{figure}

When a sample attached to the port [3] the amplitude of the interference term is   $2\gamma^2\Gamma V_cV_{in}$, where $\Gamma$ is the  coefficient describing the microwave reflection from the sample.  One can see that the ratio between this interference term and the amplitude of the interference term corresponding to Z=0 yields  the normalized interference term $\Gamma \cdot cos(\phi)$, which is proportional to the magnitude of the reflection coefficient from the sample. The  phase $\phi$ contains a  contribution related to the  phase of the reflection coefficient.  Thus the product $\Gamma \cdot cos(\phi)$ contains both the amplitude and the phase of the reflection coefficient. Below we will name the product as reflection coefficient.  Fig. \ref{reflection}(a) presents the reflection coefficient $\Gamma \cdot cos(\phi)$ of Sample C  as a function of the frequency. Sample C has the same structure  as Sample B.  

The reflection coefficient depends on the impedance $Z$ of the microwave circuit containing the sample.  A low frequency  approximation of the microwave circuit   is shown in Fig.\ref{simdiagram}. The impedance depends on contributions from the 50 Ohm terminal  resistor $R_T$, contact resistance $R_c$ and the sample.  At high frequencies the impedance may contains contributions related to a  geometry of the sample holder, in particular,  to finite lengths of the MW feeding leads. The change of the microwave reflection due to the superconductivity depends on the coupling between the sample and microwaves.  Stronger coupling makes larger temperature variations of the reflection coefficient near the superconducting transition. The difference between reflection coefficients in the superconducting and normal states is proportional to the superconducting current $I_s$ in the sample. The difference contains contributions related to variations of both amplitude and the phase of the reflection coefficient. Fig. \ref{reflection}(b) presents the  difference $\Delta (\Gamma \cdot cos(\phi))$ between two reflection coefficients obtained  in superconducting (T=4.6K) state and normal (T=13.6K) states as a function of the frequency.The data  show  relatively weak and quite uniform variation of the difference $\Delta (\Gamma \cdot cos(\phi))$  with the frequency  indicating a  uniform microwave coupling to the superconducting condensate at different frequencies. In addition Fig. \ref{reflection}(b) shows the temperature variation of the reflection from the  stage without the sample. The trace indicates that the temperature variations of the  reflection in the microwave setup itself are negligibly small \cite{kitano2008}.

To evaluate more quantitatively the microwave coupling   we note that the current through the system $I_\omega$ is related to the circuit impedance $Z$:  $I_\omega=V_\omega/Z$,  where $V_\omega$ is the microwave voltage applied to the circuit (port [3]).   At small impedance variations the current changes are 

\begin{equation}
\Delta I_\omega=\Delta( \frac{V_\omega}{Z}) =-\frac{\gamma V_{in}}{Z_0} \cdot \Delta \Gamma
\label{current_var}
\end{equation}   
, where we have used the relation between the impedance $Z$ and the reflection coefficient $\Gamma$: 
$Z/Z_0=(1+\Gamma)/(1-\Gamma)$ and the relation between incident calibrated voltage $\gamma V_{in}$ at port [2] and the voltage $V_\omega$ applied to the sample: $V_\omega=\gamma V_{in}(1+\Gamma)$.    $Z_0=50$ $\Omega$ is impedance of the coaxial line. 

Near the superconducting transition the impedance varies mostly due to the superconducting contributions.  To evaluate the strength of the superconducting current $I_s$ we  substitute  the temperature change of the current $\Delta I_\omega$ in Eq.(\ref{current_var}) between the normal and superconducting states by the superconducting current  $I_s$. In the case of the  "parallel" electrical connection of the superconducting layer to the end of the coaxial line the substitution is exact. For the circuit presented in Fig. \ref{simdiagram} the relation between two currents is $I_s=(1+R_c/R_N)\Delta I_\omega$, where $R_N \sim$120 Ohm is the normal resistance of the sample.  The contact resistance $R_c$ can be evaluated  from the temperature variation of the reflection coefficient $\Gamma$. Indeed if the resistance $R_c$ is large  in a comparison with the terminal resistance $R_T$ one should expect a small temperature variation of the impedance $Z$, since in both normal and superconducting states the current through  the resistance $R_c$ and the sample is small.  Fig.\ref{reflection}(a) shows that the reflected power $P_r=\Gamma^2(\nu) P_{in}$ is less than 50\%  in the studied frequency range. It indicates that the  impedance of the sample stage $Z$ is quite close to the coax impedance  $Z_0$: $Z\approx Z_0$.  Neglecting the difference one can related  small variations of the reflection coefficient with small  variations of the impedance:  
\begin{equation}
\Delta \Gamma =\frac{2Z_0\Delta Z}{(Z+Z_0)^2}\approx \Delta Z/2Z_0 
\label{dG}
\end{equation}   

For the circuit shown in Fig. \ref{simdiagram}  the temperature variation of the impedance between the normal ($R_N$) and superconducting ($R_S \sim 0$) states is   
\begin{equation}
\Delta Z=R_N \frac{R_T}{R_T+R_c} \frac{R_T}{R_T+R_c+R_N}
\label{dZ}
\end{equation}   

Eqs.(\ref{dG}) and (\ref{dZ}) yield that at $R_T=50$ Ohm and $\Delta G>0.1$, which corresponds to Fig.\ref{reflection}(b), the contact resistance $R_c$ is below 70 Ohms in the studied frequency range. Thus the relation $I_s=\Delta I_\omega$ underestimates  the actual superconducting current $I_s=(1+R_c/R_N)\Delta I_\omega$ by about 60\%. This estimation error is much smaller than the observed frequency variations of the nonlinear response and is neglected below.    

Experiments indicate that at small currents the nonlinear response is proportional to the applied power  $P$ and, therefore,  to the square of the superconducting current $I_s$.  Eq.(\ref{current_var}) leads to the following evaluation of  the nonlinear response at a small current:
 
\begin{equation}
\Delta R=\kappa I_s^2 \approx \kappa \cdot (\gamma\Delta \Gamma)^2 \cdot \frac{P_{in}}{Z_0}
\label{resistance_var}
\end{equation}   
, where $\kappa(\nu)$ is a frequency dependent nonlinear coefficient and  $P_{in}=V_{in}^2/Z_0$ is the calibrated power of the incident radiation at port [2]. 

\begin{figure}[t]
\includegraphics[width=3.4in]{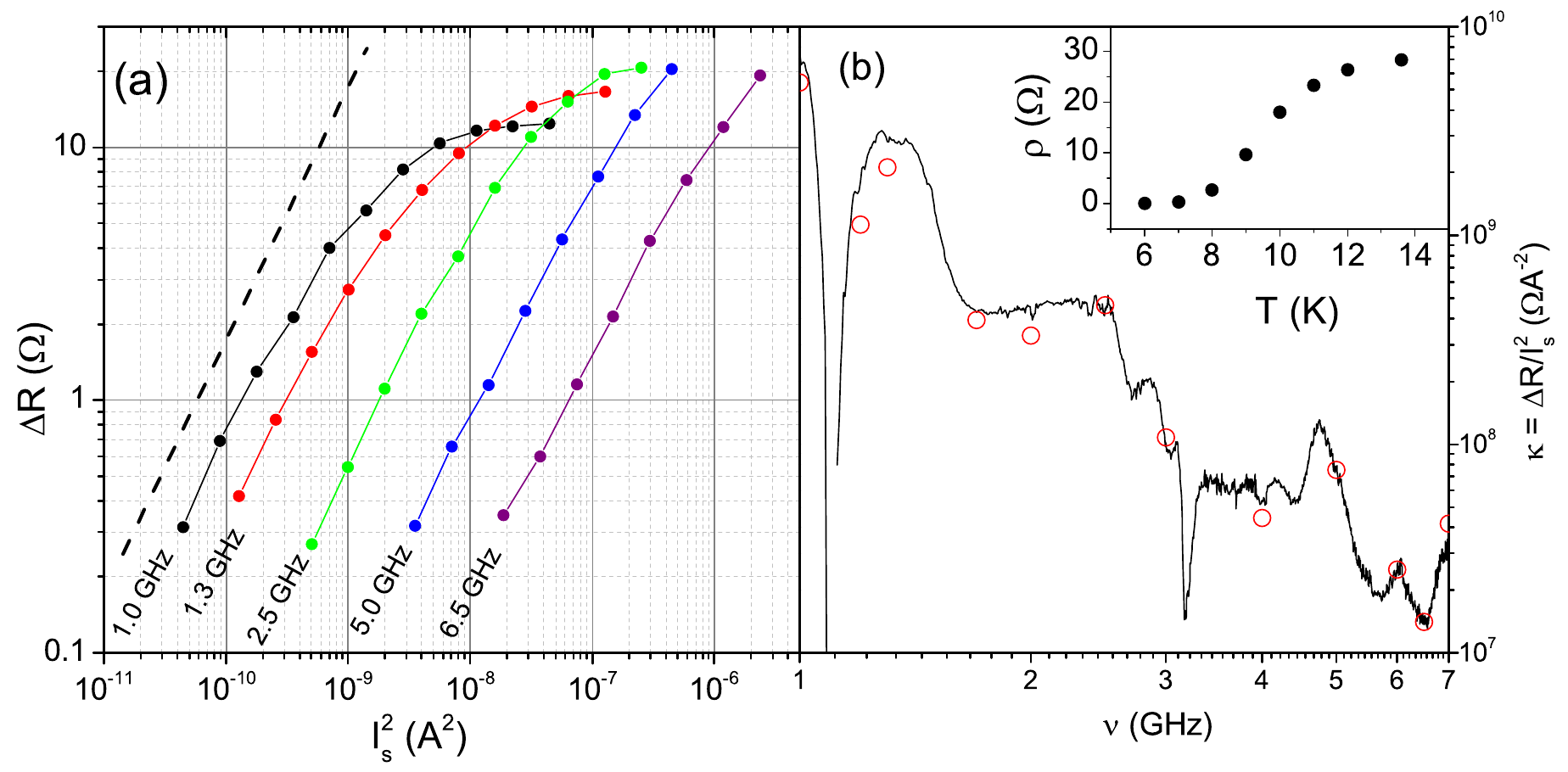}
\caption{(a)Dependence of the microwave induced variation of the resistance $\Delta R$ on the square of microwave current, which was estimated in accordance with Eq.(\ref{resistance_var}) at different frequencies as labeled; (b) Frequency dependence of the nonlinear coefficient $\kappa(\nu)$=$\Delta R/I_s^2$ obtained in the limit of small MW currents and $dc$   biases. Solid line  presents a continuous frequency sweep.  Red open circles are obtained from  the dependencies presented in Fig.\ref{atten}(a) using Eq.(\ref{resistance_var}). All dependencies are obtained at T=8K. The insert shows the temperature dependence of the sheet resistance per square of Sample C.}
\label{atten}
\end{figure}

Fig.\ref{atten}(a) presents the dependence of the  microwave induced change of the resistance on the square of the superconducting  current at different frequencies as labeled. The response, which is similar to the one obtained on  other samples, shows the strong decrease in the magnitude at high driving frequencies.  Fig.\ref{atten}(b) presents the frequency dependence of the nonlinear coefficient $\kappa(\nu)$=$\Delta R/I_s^2$. The dependence is obtained in the regime of a small power $P_{in}$ at which the nonlinear response is proportional to the square of the superconducting current $I_s$. The figure demonstrates  again the strong reduction  of the  response to the applied microwave current at high driving frequencies.  

At low frequency $\nu \sim$1 GHz the obtained coefficient of nonlinearity $k(\nu)$ is in accord with the one obtained in $dc$ domain experiments in a similar sample (sample S2 in \cite{bo2013}). The sample S2 has a slightly higher ($\delta T_c \sim$1 K) transition temperature.  In the $dc$ experiments the  nonlinear coefficient was obtained from the  relation $E=\rho_0 J_{dc}+\gamma_{NL} J_{dc}^3$ between  the electric field $E$ and applied current density $J_{dc}=I/W$, where $W=200 \mu$ is the width of  samples and $\gamma_{NL}$ is the nonlinear coefficient. The relation can be rewritten as $V=EL=R_0\cdot I+\gamma_{NL} L(I/W)^3$ yielding  $\kappa = 3\gamma_{NL} L/W^3$, where $L=800 \mu$ is the length of  samples. At T=9K (corresponding to T=8K for Sample C) the nonlinear coefficient $\gamma_{NL} \approx$5 $\Omega m^2/A^2$ (see Fig.3a in \cite{bo2013}). It yields $\kappa_{S2}$=1.5$\cdot$10$^9$ $\Omega/A^2$. This value is in a quantitative agreement with the one shown in Fig.\ref{atten}b at low frequencies.  The correspondence between the nonlinear coefficients obtained by different experimental methods demonstrates that the reflection method provide a reasonable evaluation of the applied microwave current.       

In conclusion, the  presented reflection measurements indicate that the microwave radiation is quite uniformly coupled to  the superconducting electrons at different frequencies.  Analyzed in terms of the  applied superconducting current the nonlinear response demonstrates significant reduction at high driving frequencies.

\end{document}